\newcommand{\beq}{\begin{equation}}
\newcommand{\eeq}{\end{equation}}

\newcommand{\bear}{\begin{eqnarray}}
\newcommand{\eear}{\end{eqnarray}}

\newcommand{\tn}{\textnormal}
\documentclass{ws-procs9x6}

\begin{document}

\title{The NEXT experiment. Towards phase I}

\author{D. Gonzalez-Diaz$^*$ for the NEXT collaboration}

\address{Laboratorio de F\'isica Nuclear y Astropart\'iculas, Universidad de Zaragoza\\
Calle Pedro Cerbuna 12, 50009 Zaragoza, Spain\\
\address{Laboratorio Subterr\'aneo de Canfranc\\
Paseo de los Ayerbe s/n, 22880 Canfranc Estaci\'on, Huesca, Spain\\}
$^*$E-mail: diegogon@unizar.es\\
http://www.unizar.es}

\begin{abstract}
Phase I of the NEXT-100 $0\nu\beta\beta$  experiment (NEW) is scheduled for data taking in 2015
at Laboratorio Subterr\'aneo de Canfranc in the Spanish Pyrenees. Thanks to the light proportional technique, NEW anticipates an outstanding energy resolution nearing the Fano factor
in Xenon (0.5-1\%FWHM@$Q_{\beta\beta,^{136}Xe}$), with a TPC-design that allows tracking and identification of the double end-blob feature of the $0\nu\beta\beta$ decay. When properly mastered, the combination of these two assets can suppress the irreducible $2\nu\beta\beta$ and (single-blob) $\gamma$ backgrounds from natural radioactivity to minute levels, of the order of $5\times{10^{-4}}$\,ckky. Given our knowledge of the available phase-space as obtained from neutrino oscillation experiments, this feat will expectedly allow for a sensitivity to the effective electron neutrino mass of $m_{\beta\beta}\simeq 30$\,meV for exposures at the $20\,\tn{ton}\times$\,year scale. Hence, ultimately, a full survey of the inverse hierarchy of the neutrino mass ordering appears to be within reach for a ton-scale experiment based on this technology.

NEW, with 10\,kg of Xenon 90\%-enriched in $^{136}$Xe, sets an unprecedented scale for gaseous Xenon TPCs and will be an important milestone for its anticipated upgrades (100\,kg and 1\,ton). I briefly summarize the status of the NEXT experiment, from the main results obtained with $\sim\!1$\,kg prototypes that substantiate the concept, to the ongoing works for deploying its first phase.
\end{abstract}


\bodymatter

\section{Introduction}\label{aba:sec1}

The controversial observation in 2001 of the $0\nu\beta\beta$ decay \cite{KK} has made the case for independent confirmation in different isotopes increasingly compelling \cite{Avignone:2007fu,Cadenas_Rev_bb}. Indeed, great success has been recently achieved through the combined (negative) searches in $^{136}$Xe (EXO \cite{EXO}, KamLAND-Zen \cite{KAM}) and $^{76}$Ge (GERDA \cite{GERDA}) isotopes, strongly increasing the tension with the previous finding\cite{KK}, to the extent of making it statistically incompatible to at least 90\% confidence level. Thus, re-interpreting the new results either as positive claims or as larger neutrino masses than the reported bounds seemingly requires of large unknown systematic effects plaguing the new experiments as well as the `state of the art' theoretical calculations.

Complementary to these ongoing searches, background suppression through some novel and intriguing proposals is being explored. Taking $^{136}$Xe as an example, the Barium tagging \cite{BaTag} and the double end-blob identification \cite{refTDR} clearly stand out as potential, highly unambiguous, `smoking guns'. The second idea, pursued by the NEXT collaboration following earlier ground-breaking work \cite{GroundBreakXe}, has been recently demonstrated in $\sim\!1$\,kg-scale TPCs \cite{NEXT_LBNL,NEXT_DEMO2}. The bottom-line of NEXT is the identification of the double end-blob feature of the 2e- tracks produced in a $\beta\beta$ decay, while being able to energetically resolve the $0\nu\beta\beta$ and $2\nu\beta\beta$ modes.

  \begin{figure}[h]
 \centering
 \includegraphics*[width=4.8cm]{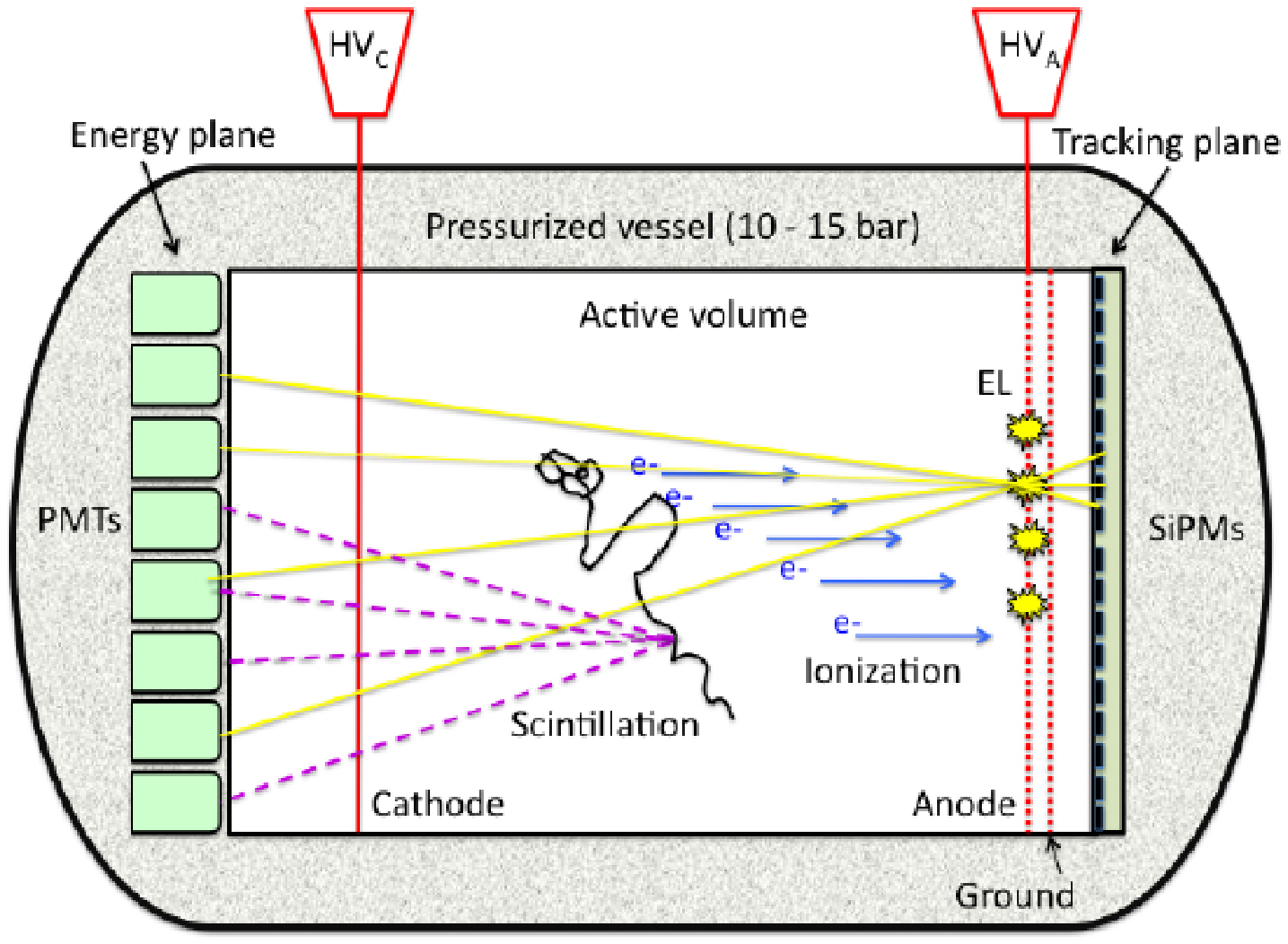}
 \includegraphics*[width=4.1cm]{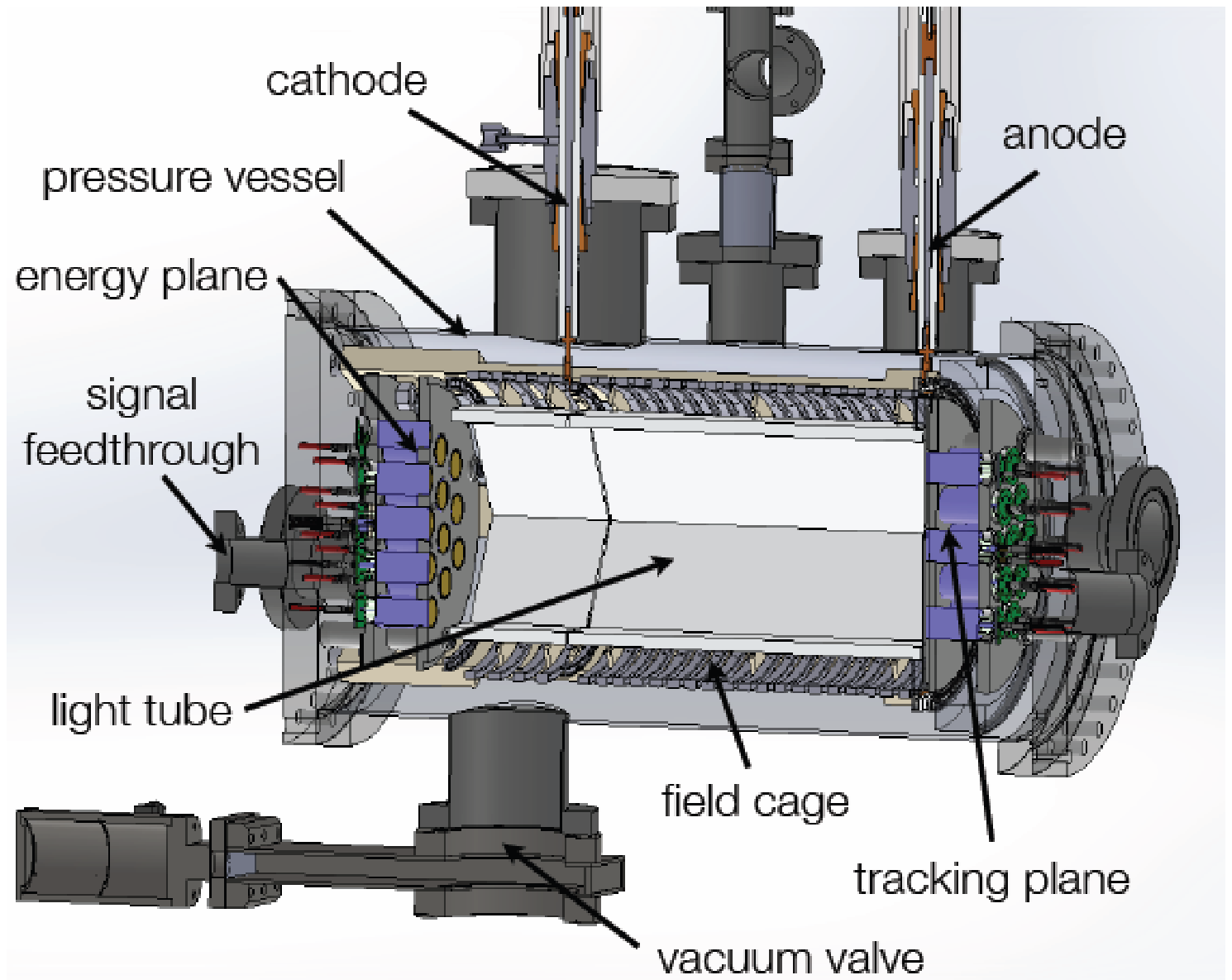}

  \includegraphics*[width=9cm]{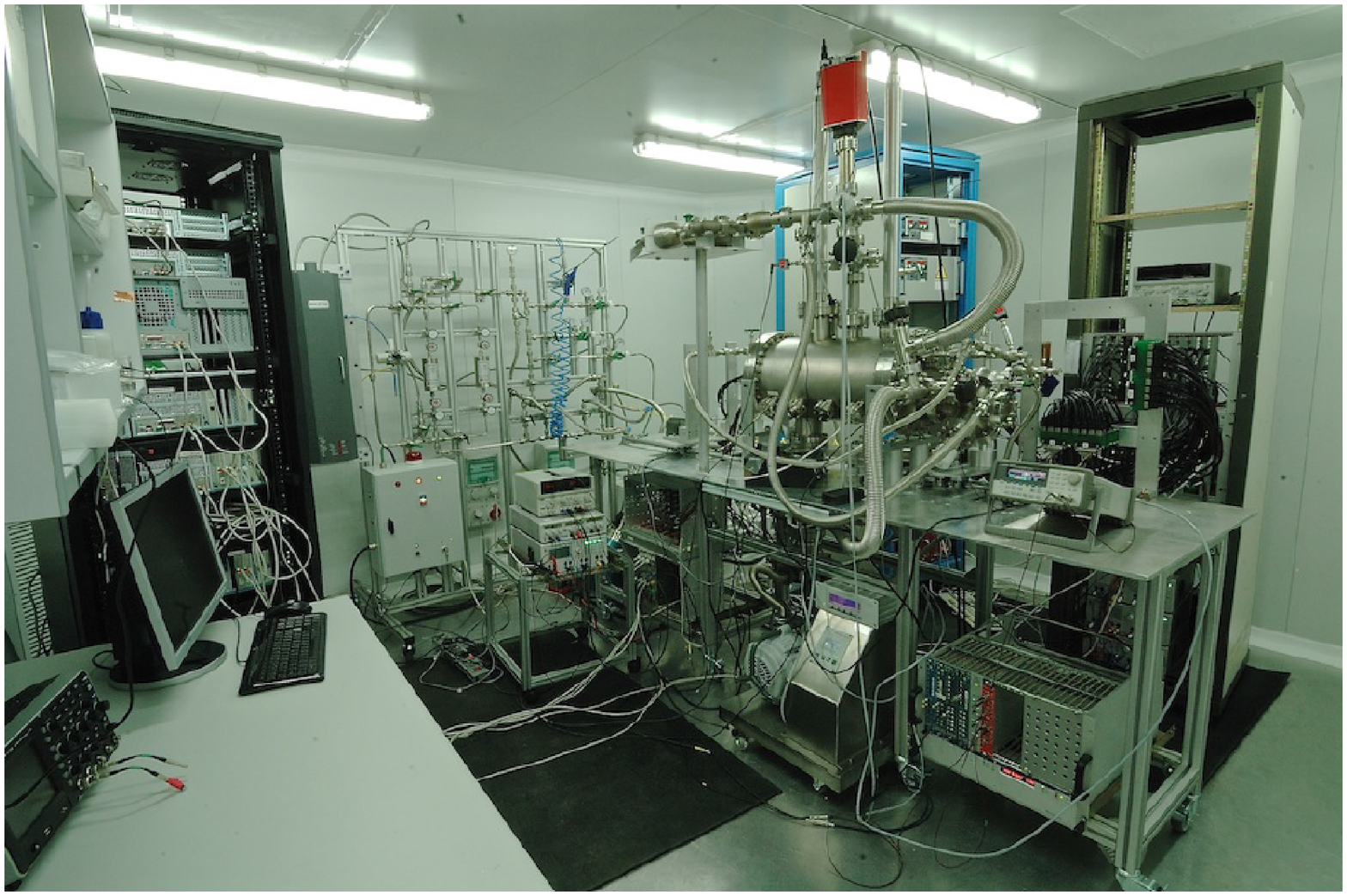}
 \caption{Up-left: conceptual design of the technological baseline of the NEXT experiment. Up-right: technical design of the NEXT demonstrator. Down: photo of the NEXT demonstrator in its room at IFIC, Valencia.}
 \label{design}
 \end{figure}

Thanks to the light-proportional technique (i.e, electroluminescence), the NEXT technological demonstrator (Fig.\ref{design}) has achieved an energy resolution of 1.82\%FWHM for 0.511\,MeV electron tracks, enough for completely separating the two $\beta\beta$ modes at the $Q_{\beta\beta,^{136}Xe}=2.45$\,MeV scale (0.83\%FWHM). On the other hand, with the help of an optimized tracking plane, the probability of mis-identifying a second blob at the tracks tail was seen to be as low as $10^{-3}$.
The first stage of the experiment (NEW) is currently under construction, scheduled for data taking in year 2015. With its 10\,kg of mass, it will be able to quantify the strength of the topological signature in realistic conditions by using a nature-provided sample of the extremely rare $2\nu\beta\beta$ mode. 

\section{Performance of the NEXT technological demonstrator}

\begin{figure}[h]
 \centering
 \includegraphics*[width=\linewidth]{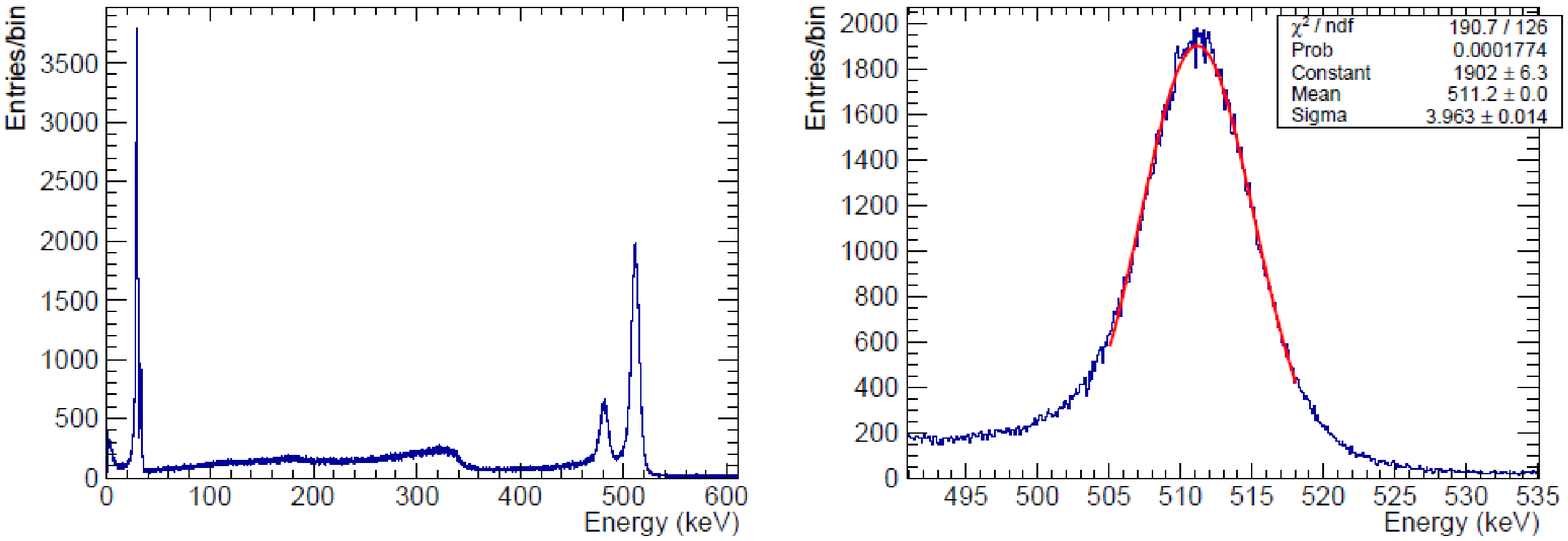}
 \includegraphics*[width=\linewidth]{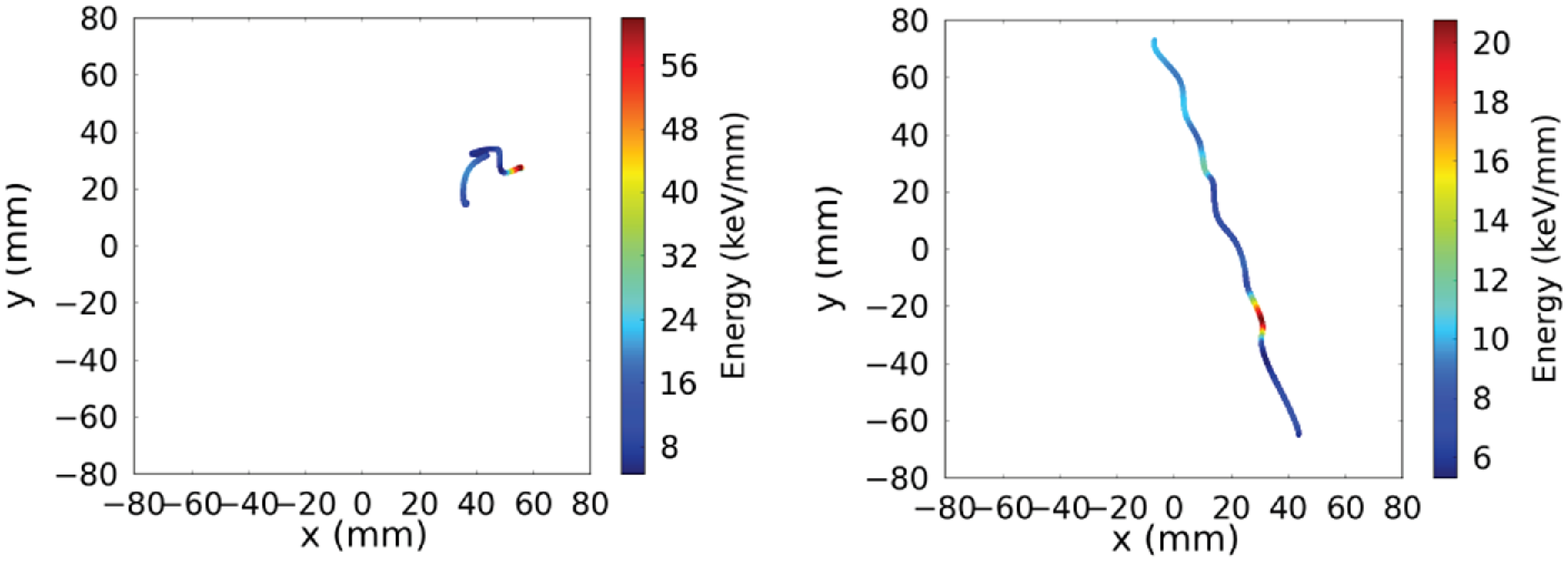}
 \caption{Up: energy resolution of the NEXT demonstrator for 0.511\,MeV tracks after fiducialization. Down: reconstructed example of the topological signature (end-blob) for a contained electron and a feature-less punching-through muon.}
 \label{ERes}
 \end{figure}

NEXT-DEMO, a 30\,cm-drift high purity 10\,bar Xe-TPC deployed at IFIC-Valencia, implemented for the first time the technological workhorse of the NEXT experiment through a separate topology/calorimetry function with two optimized planes (Fig.\ref{design}-up). The `energy plane' (performing the calorimetric measurement) was based on 19 1-inch pressure-resistant R7378A PMTs, while the `tracking plane' (performing the topological measurement) relied on a matrix of 1\,cm-pitch TPB-coated S10362-11-050P SiPMs covering 256\,cm$^2$. The inner part of the field cage was covered with TPB-coated PTFE panels, to improve light collection. The TPC was hence fully read in light mode, with the primary ionization converted to light through the electroluminescence created during the electron transit in between two 5\,mm-distant grids, at a field around 2\,kV/cm/bar. Besides reconstructing the end-blob feature (Fig. \ref{ERes}-down), the information obtained from the tracking plane allowed correcting x-y distortions of the light collection efficiency at the energy plane, providing an outstanding energy resolution of 1.82\%FWHM for 0.511\,MeV electron tracks (Fig. \ref{ERes}-up). The chosen fiducialization (removing tracks originated close to the chamber boundaries) extrapolates to 90\% of the active volume in NEXT-100.

\section{Other prototypes}
The NEXT-DBDM TPC \cite{NEXT_LBNL}, built at Lawrence Berkeley National Lab, achieved an energy resolution of 1\%FWHM for 0.662\,MeV $\gamma$'s on a 10\,bar, $\sim$1\,kg chamber, in its very central axial region (less than a factor $\times 2$ shy of the intrinsic Fano limit). The absence of tracking plane at that stage impeded a dedicated study of the energy resolution over the full chamber, but it showed what could be done. With its new tracking plane installed, NEXT-DBDM is currently employed for studying nuclear recoils in the context of dark matter detection, a subject of intense R\&D in the collaboration.

NEXT-MM, developed at Zaragoza University \cite{NEXT-MM_first}, is a 70\,l TPC currently running in the 1-2.7\,bar pressure range under a novel microbulk-Micromegas readout working in charge-mode. Its long drift region (38\,cm) and finely $0.8$\,cm$\times0.8$\,cm segmented $700$\,cm$^2$ readout makes it very well suited for the study of the electron swarm properties of some enterprising Xe-based mixtures. Exemplarily, drift velocity, diffusion and attachment coefficients have being recently extracted for the first time for Xe-Trimethylamine(TMA) mixtures \cite{Diego_NEXT}. NEXT-MM is currently undergoing a 10\,bar commissioning run with this mixture.


\section{NEXT-stage I (NEW)}

\begin{figure}[t!]
 \centering
 \includegraphics*[width=\linewidth]{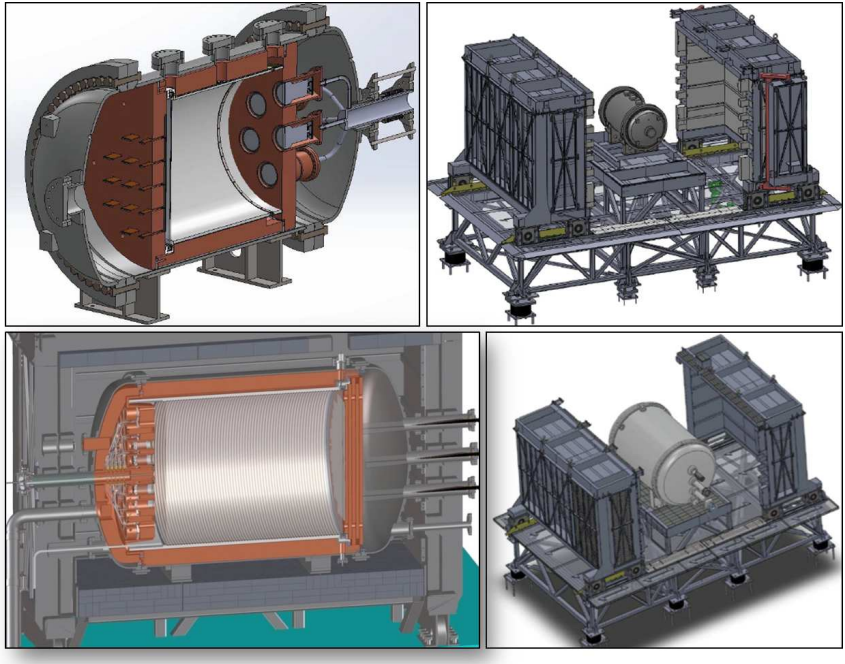}
 \caption{Up: design of NEXT stage 1 (NEW) and NEW inside the lead castle, due for construction in 2014. Down: similarly, NEXT-100 and castle.}
 \label{NEXT_AND_NEW}
 \end{figure}

\begin{figure}[h!]
 \centering
 \includegraphics*[width=11.5cm]{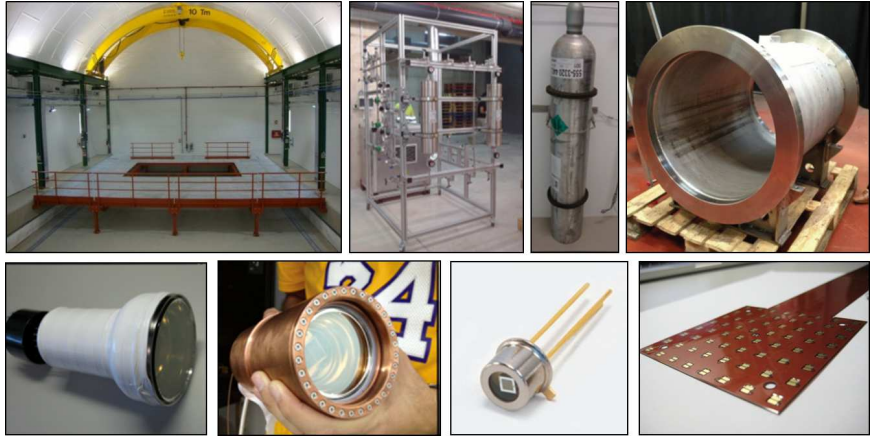}
 \caption{Some relevant hardware components of NEW. From left-right, up-down: the platform supporting the experiment at Laboratorio Subterr\'aneo de Canfranc (LSC), the gas system, 100\,kg $^{136}$Xe bottle, the vessel body, R11410-10 Hamamatsu PMT, PMT copper can, S10362-11-050P SiPM, piggy-tail SiPM board.}
 \label{NEW_ALL}
 \end{figure}

The collaboration is currently immersed in the construction of the first stage of NEXT by end of 2014, with data taking starting in 2015. It is conceived as a 2:1 down-scaled version of NEXT-100, implying readout planes of about 20\% the size of the final ones, with 10-15\,kg $^{136}$Xe mass (10-15\,bar) and being based on the final technological solutions, as much as possible. A comparison between the first stage of NEXT and NEXT-100 is illustrated in Fig. \ref{NEXT_AND_NEW} (up row: NEW, low row: NEXT-100), with the lead castle for shielding shown on top of a seismic platform on the right column. Some relevant innovations as compared to the demonstrator are the introduction of radio-pure piggy-tail kapton PCBs and R11410-10 PMTs (Fig. \ref{NEW_ALL}), enclosed in pressure-resistant vacuum Cu cans, as well as the inner copper shielding and several additional service ports. Importantly, most components of NEW have been already screened for radio-purity and selected as part of the NEXT-100 screening campaign \cite{RadioPurity}.

\begin{figure}[h!]
 \centering
 \includegraphics*[width=6.7cm]{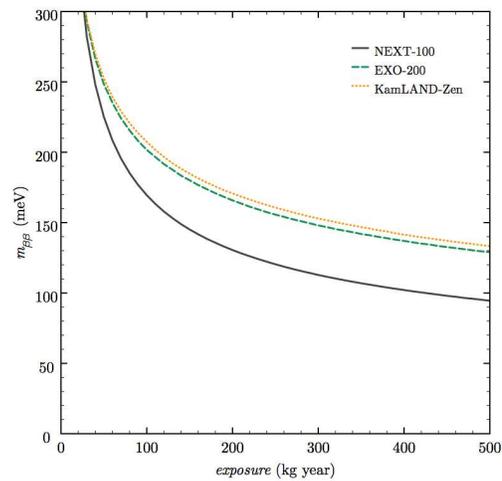}
 \caption{Comparison, mainly for illustration, of the sensitivity to the effective Majorana mass as a function of exposure for next generation Xenon experiments.\cite{CadenasLAST} EXO-200 and KamLAND-Zen currently set the best sensitivity limits so far published.}
 \label{NEXTfut}
 \end{figure}

It has been argued \cite{CadenasLAST} that, despite its late start with respect to already
running $0\nu\beta\beta$ experiments, NEXT-100 shows a competitive figure as a function of exposure, in virtue of its ultra-low background and outstanding energy resolution (Fig.\ref{NEXTfut}).

\newpage
\section{Conclusions}
The $0\nu\beta\beta$ quest anticipates an inspiring landscape for years to come, gaining steam from complementary $\nu$-oscillation measurements, direct $\nu$-mass determination, and cosmological constraints. A Majorana nature for the neutrino is strongly favoured by theoretical considerations, while future $\nu$-oscillation experiments may tell us whether the mass ordering chosen by nature is inverse. It is then most certain that the NEXT technique will be able to contribute to the long-awaited discovery of the $0\nu\beta\beta$ decay mode, either providing a first measurement or a highly unmistakable confirmation by resorting to its unique double end-blob identification capabilities.

\end{document}